# Multi-parameter sensing enabled by modular fibre-optic assemblies

Kenny Hey Tow, Joao M B Pereira, Miguel Soriano-Amat, Markus Persson, Kristian Angele, Mats Billstein, María R. Fernández-Ruiz, Camilo Escobar-Vera, Åsa Claesson

*Abstract*— **Industry is transitioning from manually monitored components and processes to data-driven solutions. At the heart of this transformation is predictive maintenance, which relies on simultaneous, real-time monitoring of key operational parameters such as temperature and vibration to anticipate and prevent equipment failures. In this work, we present a modular approach to fibre-optic sensing, where different types of optical fibres and other wires are combined to compact, hybrid cable assemblies, customized for each application. These fibre-optic assemblies can be embedded or integrated in various settings, enabling multi-parameter sensing and the measurement of new parameters.**

## I. INTRODUCTION

Industry 4.0 is driving a paradigm shift in how industries manage, monitor, and maintain their physical assets. With the integration of Industrial Internet of Things, and advanced analytics, maintenance strategies are evolving from reactive and preventive approaches to more intelligent, predictive maintenance models. A critical enabler of this shift is the ability to continuously and accurately monitor key operational parameters—particularly temperature and vibration—which are well-established indicators of equipment health. Abnormal vibrations are often early signs of mechanical issues such as misalignment, imbalance, or bearing defects, while temperature anomalies can indicate problems such as overheating, lubrication failure, or electrical insulation breakdown [1]. Detecting such changes early allows taking corrective actions before catastrophic failures occur, reducing downtime, repair costs, and safety risks.
Fibre-optic sensors are particularly adapted for predictive maintenance. Compared to traditional sensors, they are less invasive, easier to install in assets, and a single optical fibre can be used to monitor sensing parameters such as temperature, strain and vibration with high spatial resolution over large areas (kilometer ranges) in harsh environments. For instance, optical fibre sensor technologies such as Fiber Bragg Gratings (FBGs), Distributed Acoustic/Temperature Sensing (DAS/DTS) enable temperature and vibration to be measured along a single optical fibre with high spatial and temporal resolution. The latter technology has mainly been developed for applications such as long-range pipeline and energy transport, and thick ruggedized, field-deployable optical fibre cables are used— typically 5 mm to 1 cm thickness, and 3 to 4 mm for slimmer and more flexible cables [2].

The thickness of sensing cables can sometimes limit their applicability in specific scenarios. In industries such as aerospace and automotive, embedding fibre-optic cables into composite materials for structural health monitoring demands extremely thin and flexible fibres (< 1 mm) to not compromise the structural integrity, asset flexibility and adding unwanted weight to composite structures [3] and in fabrics for smart textile [4]. Similarly, fibre-optic sensing cables can pose significant challenges during installation in narrow, confined spaces such as bends, elbows, etc. [5], limiting their deployment in monitoring campaigns.

Instead, individual optical fibres are used but they are often too fragile for embedding or deploying in harsh conditions, resulting in fibre breaks and very low sensor lifetime [6-9]. Moreover, one fibre is usually needed per parameter monitored, which complicates and limits their installation and deployment. One can use the coating, for instance, the electrical conductivity of carbon or nanocomposite coated fibres, in multi-sensing systems in association with distributed optical fibre sensing [10], or simultaneous environmental and temperature measurements in quasi-distributed sensing schemes [11], provided one can have access to the fibre coating during the measurement.

As an alternative, we propose a modular approach to fibre-optic sensing, in which sensors of different types and for

This work was supported by Sweden's Strategic Innovation Program for Smarter Electronic Systems (2022-00835, 2023-03072), funded by Sweden's Innovation Agency Vinnova, the Swedish Energy Agency, and Formas and in part by MCIN/AEI/10.13039/501100011033 under Grant SEASNAKE+, ref: PCI2023-145978-2 of the CET Partnership 2022 joint call. The work of M.R.F-R. was supported by MCIN/AEI/10.13039/501100011033 and European Union «NextGenerationEU»/PRTR under grant RYC2021-032167-I. This work was supported in part by the Spanish MCIN/AEI/10.13039/501100011033, and FEDER Una manera de hacer Europa under Grant MOTION: ref. PID2022-140963OA-I00.

Kenny Hey Tow, Joao Pereira, Miguel Soriano-Amat, Markus Persson and Åsa Claesson are with RISE Research Institutes of Sweden, RISE Fiberlab, Fibervägen 2-6, 824 50 Hudiksvall, Sweden (e-mail: kenny.heytow@ri.se).
Kristian Angele and Mats Billstein are with Vattenfall AB, R&D Laboratories, Älvkarleby, 814 70, Sweden
María R. Fernández-Ruiz and Camilo Escobar-Vera are with the Universidad de Alcalá, Escuela Politécnica Superior, Sensors and Photonic Technologies, UAH, Associate Unit to CSIC by Institute of Optics, 28805 Madrid, Spain.

different measurands can be combined to miniaturized packages and embedded or integrated in various settings. In this scenario, kilometers long, compact (< 1mm thick) hybrid optical cable – referred to hybrid fibre-optic assemblies [12] – combining Fibre Bragg Gratings, optical fibres for distributed sensing and thin electrical wires in a thin bundle. When embedded in a host material, this fibre-optic assembly enables multi-parameter sensing during manufacturing, testing, and throughout the operational lifetime of the asset, improving predictive maintenance capabilities and providing feedback to the design and the manufacturing process.

In this paper, we describe the fabrication process of these fibre-optic assemblies in Section II. Using this technique, kilometers of fibre-optic assemblies can be custom-made for desired applications. They can be directly embedded into assets or deployed in harsh environments. In Section III, we describe three use cases, in which these fibre-optic assemblies were used, highlighting their versatility for deployment and their added benefits for monitoring purposes. The relevance of developing these fibres and future applications with these fibres are discussed in the last section.

## II. Fibre-optic assembly fabrication

A modular process for manufacturing fibre-optic assemblies was developed in our fibre drawing facilities using a conventional fibre-optic extruder line. All the optical fibres can be fed into the extruder head in one setup, and the entire assembly is buffered in one step, with a single outer buffer coating. For careful control of wire tension and position in the assembly, each fibre has a separate payoff. When relevant, SZ-rotation of the fibres can be introduced. Hybrid configurations with one or more non-optical wires (e.g. electrical wires or aramid strength elements) can be manufactured using the same extruding process.

The fibre coatings are kept intact during the cable manufacturing, which makes the assemblies durable and easy to connectorise. The buffer material can be selected from a range of standard fibre-optic buffering materials, such as Nylon, ETFE, PVC, Hytrel, PEEK, or PFA. The outer diameter of the miniature cable can be anywhere from around 100µm (using ultrathin fibers [13]) to several millimeters, depending on the complexity of the assembly, the number of fibres/wires inside, and the requirements of the use case.

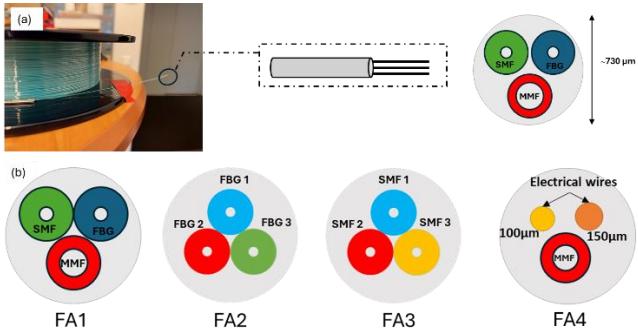

Fig. 1. (a) Example of a spool of fibre-optic assembly produced at RISE's fibre manufacturing facilities. (b) Sideview illustrations of four fibre assemblies made up of different types of optical fibres (FA1), three Fibre Bragg Grating arrays (FA2), three single-mode fibres (FA3) and a hybrid assembly comprising of 1 multimode optical fibre and 2 electrical wires (FA4) extruded together.

This modular approach makes the design options for these miniature cables almost endless. The optical fibre types can be chosen freely and can have different coatings and dimensions. This is particularly useful for multi-parameter sensing since one can have a set of different optical fibres to measure several parameters simultaneously. For instance, one can co-extrude FBG arrays with telecom optical fibres for multi-sensing applications, or combine various fibres for sensing, power delivery and communication. Hybrid assemblies with electrical wires co-extruded with optical fibres have also been demonstrated. The fibres/wires inside the assemblies are easily separated and connectorized by removing the buffer material. From an optical perspective, it is important to consider microbend losses (especially if the fibres have thin and hard coatings such as polyimide), and packaging-induced strain of for e.g., FBGs.

RISE has been manufacturing these fibre-optic assemblies over these past few years for dedicated applications. In this communication, we used four of such fibre-optic assemblies that were custom-made for the use cases described in the next section. They are all composed of three fibres/wires, co-extruded over hundreds of metres as shown in Fig 1(a). The composition of these different fibre-optic assemblies are schematically represented in Fig. 1(b): (i) FA1, made up of three different optical fibres (1 multimode fibre, 1 single mode fibre and 1 FBG array), (ii) FA2 and FA3, composed of three similar fibres, Fibre Bragg Grating arrays and single-mode fibres respectively, and (iii) FA4, a hybrid assembly with 2 different electrical wires and a multimode fibre extruded together.

## III. Use cases: examples of the use of fibre-optic assemblies for various sensing applications

For multi-parameter sensing, different types of optical fibres are often used, which comes with its own challenges during installation. The use of one miniature cable that contains all the required optical fibres can reduce the complexity of embedding in assets and deployment during field measurements. Moreover, an astute choice of buffer material for the fibre-optic assembly can bring additional protection to the optical fibres against harsh environment and chemicals. In this section, we present three real use cases in which the use of fibre-optic assemblies enhanced the monitoring process either by allowing multi-sensing functionality or by unlocking the measurement of new sensing parameters.

### A. Use Case 1: Multi-parameter sensing using 3 different co-extruded optical fibres (FA1)

Monitoring temperature and vibration in fluids such as oil, along with other relevant parameters, is a key component of a robust predictive maintenance strategy. It enables early detection of failures, reduces downtime, and optimizes maintenance schedules, leading to significant cost savings and increased operational efficiency. A typical use case would be distributed temperature and vibration measurement in large vessels and tanks. For instance, corrosion and early leakages can be inferred from thermal anomalies and identifying unique acoustic signatures before visible failure [14].

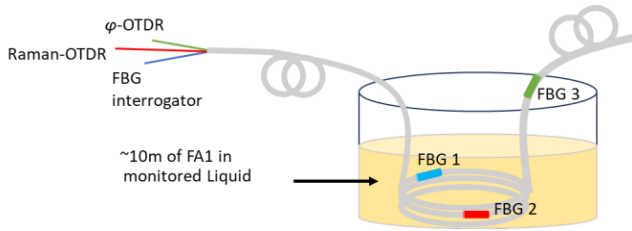

Fig. 2. Experimental scheme used to test multi-parameter sensing using FA1. The multimode fibre was probed by a Raman-OTDR to monitor temperature changes while the FBG array and the single-mode fibre were used for quasi- and fully-distributed vibration sensing.

FA1 assembly was specifically designed and fabricated for this application. It is composed of three different types of optical fibres: (i) one standard multi-mode fibre (MMF), used exclusively for distributed temperature sensing over these 70 metres; (ii) one standard telecom single-mode fibre (SMF), that can be used to measure different parameters such as strain, vibration and temperature in a distributed way and (iii) one array of 3 FBG sensors, written at lengths 22 (FBG 1), 26 (FBG 2), and 30m (FBG 3) along the FBG fibre array.

A 10-metre segment of the FA1 cable was attached at the bottom of a container filled with the monitored liquid and each of the fibres in FA1 connected to corresponding fibre-optic interrogation systems, as depicted in Fig. 2 for continuous, real-time monitoring of thermal and acoustic signals inside a liquid-filled container over a long period of time. Distributed temperature measurements obtained on the first night of the monitoring campaign, from 16:00 until 07:00 the other day, are plotted on Fig. 3(a). During this test phase, the liquid was heated to approximately 35ºC and the cooling phase of the latter monitored (one measurement every 2 minutes) by connecting the MMF of FA1 to a commercial Raman-OTDR system. A spatial resolution of 0.5 metres was used for the measurement. The FBG array and the SMF were both probed at 1 kHz using commercial FBG interrogator and phase-sensitive OTDR systems.

At around 18:30, a weak acoustic signal of 430 Hz was generated inside the liquid using an audio transducer, attached to the outside of the container. This acoustic signal was picked up by the phase-sensitive OTDR interrogator, as shown in Fig. 3(b), and the two FBGs (FBG 1 and 2) located inside the container. The corresponding frequency components obtained from the FBG measurements and phase-sensitive OTDR traces are pictured in Fig. 3 (c-d), confirming the feasibility of using FA1 for multi-parameter sensing.

*B. Use Case 2: Towards shape sensing using 3 extruded FBGs (FA2) and single-mode fibres (FA3)*

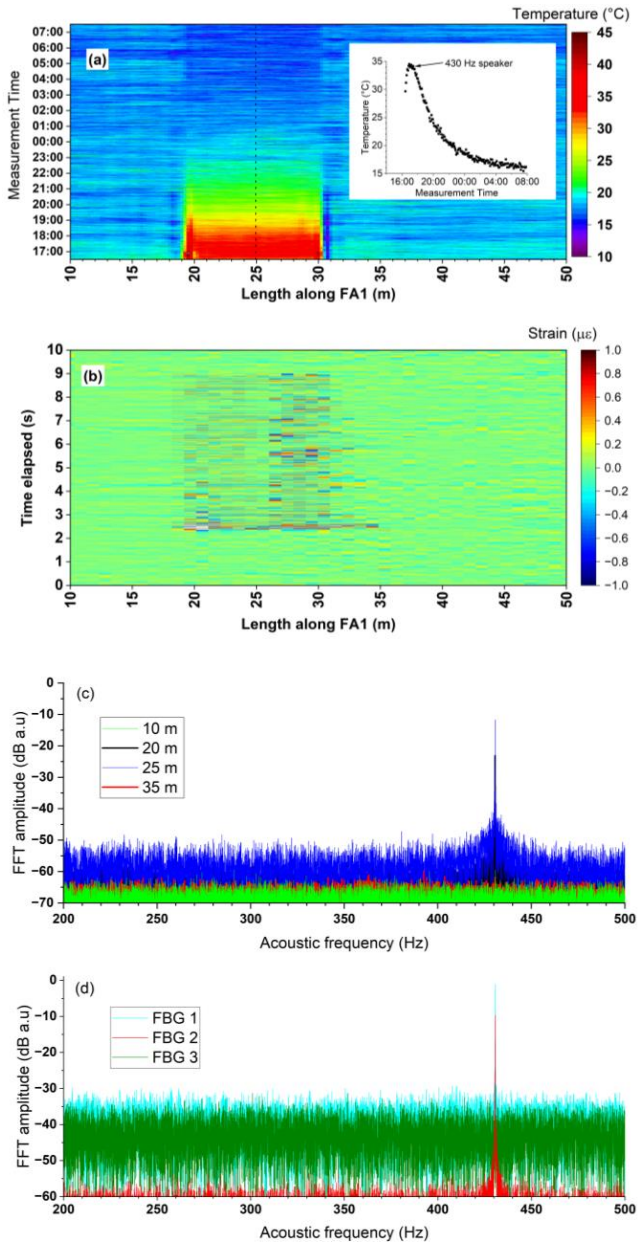

Fig. 3. (a) Distributed temperature measurements over 1 night monitoring during the cooling phase of the liquid (Inset: temperature evolution with time at position 25 metres in FA1). (b) A 10 second measurement showing the strain change induced along FA1 when a short acoustic signal was applied on one side of the container. (c) Frequency content of this acoustic signal, picked up by the part of the single-mode fibre and (d) the two FBGs (1, 2) immersed in the liquid under test.

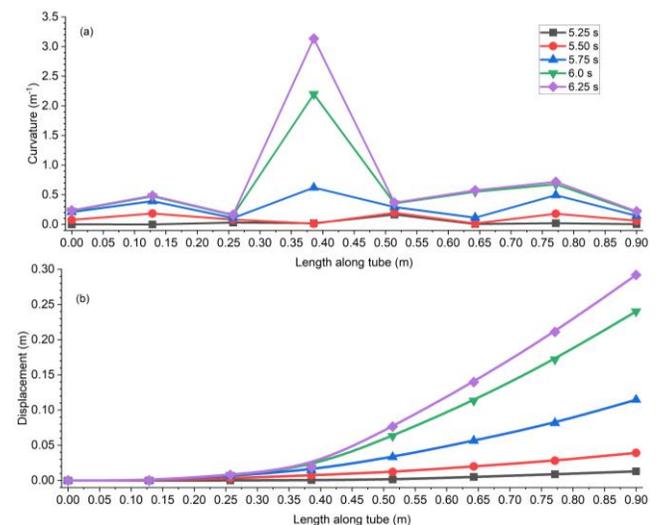

Fig. 4. (a) Curvature measurement at different positions along FA2 and (b) corresponding deflection of the beam, obtained using these curvature values.

Multicore optical fibres or multiple single-core fibres are often used for shape reconstruction. The latter alternative is particularly advantageous in terms of curvature sensitivity since a larger spacing between the fibre cores can be achieved if individual fibres are used, but it can be more complex to deploy in assets [15] since it involves placing the fibres closely together over the entire asset. The fibre-optic assembly method described in this paper makes the deployment of such multi-fibre shape sensing solution much easier.

FA2 was fabricated by co-extruding nylon with three fibre FBG arrays, each FBG being spaced by 10 centimeters. Special efforts were made during the extrusion process to ensure that the 3 arrays were minimally twisted and that the FBGs were co-located along the fibre assembly. FA2 was inserted into a steel tube to monitor the tube's shape when bent. For each bent position, the local strain mapping along each FBG arrays is measured and the corresponding curvature (Fig. 4(a)) is determined at these sensing positions [16]. Using this information, the deflection of the bent metal tube could be reconstructed with high fidelity as shown in Fig. 4(b).

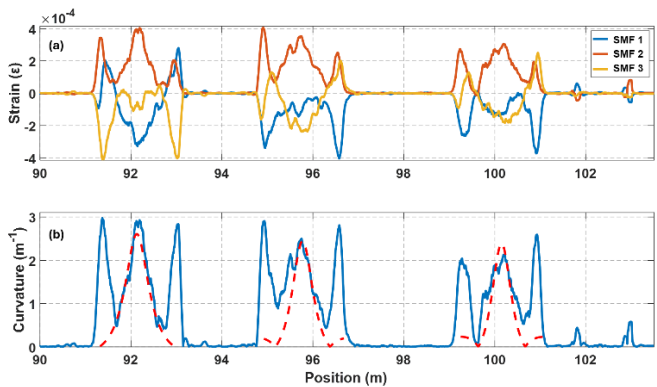

Fig. 5. (a) Strain profiles of each fibre in FA3 assembly and (b) reconstructed curvature profiles at the three bent sections. The expected curvature change, obtained from simulated catenary, is included in dotted red line

The same concept can be extended to fully distributed shape sensing measurements by using three single-mode fibres in the fibre-optic assembly (FA3) and associating it to optical reflectometry, either in frequency [17] or in time domain [18]. To illustrate this concept, three sections of FA3 between two fastened ends were bent to form a catenary to induce different curvatures along the fibre assembly in a controlled way. Curvature variations were induced by moving one of the fastened ends closer to the other. The three fibres were interrogated using time-expanded phase-sensitive OTDR technique as described in [19]. The corresponding distributed strain and the resulting curvature profiles, obtained from the differential strain measurements, are plotted in Fig. 5(a) and Fig. 5(b). For the three sections, the measured curvature values are in agreement with the expected curvature changes (represented in dotted lines), obtained by simulating a catenary shape using the parameters of the bent sections. This implies that fibre-optic assemblies such as FA2 and FA3 can be reliably used as an alternative to multi-core fibres for distributed curvature measurements for shape mapping.

*C. Use Case 3: Hybrid fibre assemblies for distributed measurement of water flow speed (FA4)*

Active distributed temperature sensing (DTS) [20] is a measuring technique used to monitor atmospheric and hydrogeological parameters that affect thermal conductivity, for e.g., seepage, water velocity, wind speed, etc. Typical active DTS sensing systems involve passing a current through the steel armouring of a fibre-optic cable to elevate its temperature along the whole cable and monitor the subsequent small temperature changes brought by a flowing liquid or gas. Commercial active-DTS systems typically allow monitoring over 1 km at a power rate of around 20 W/m with a temperature increase of 5 K [21].

Our compact fibre assembly concept can be extended for fabricating long lengths of hybrid fibre-optic assemblies (e.g., FA4), made up of 2 electrical wires co-extruded with a standard 50/125 multimode optical fibre for active-DTS measurements. Since the two electric wires are located very closely to the optical fibre in the assembly, one can expect higher and faster temperature changes to the multimode fibre when applying a current through the co-extruded electric wires. Moreover, this assembly fabrication method brings modularity since one can exactly select which parts of FA4 is heated when applying a current to the whole assembly. For instance, 3 hotspots of different lengths (50 cm, 1 metre and 2 metres) were made on a section of FA4 by directly contacting the metal wires over the desired hotspot sections during the fabrication process. As illustrated on Fig. 6, only these 3 sections were locally heated when a current was applied to the metal wires at one end facet of FA4. Thus, the sensing system becomes more energy-efficient as less electric power is needed to dedicated heat zones rather than the whole fibre cable, or more sensitive to external parameters since supplying the same electrical power creates a higher temperature gradient on the hotspot zones.

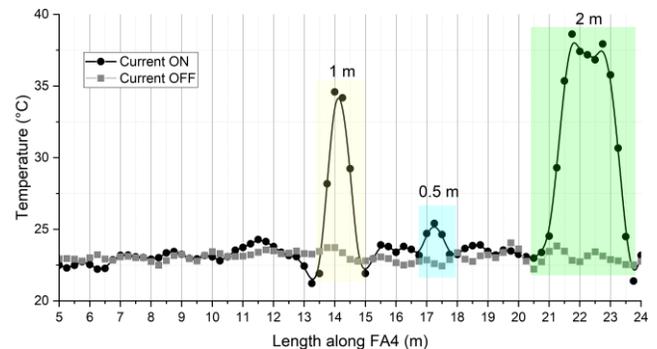

Fig. 6. Examples of three hotspots of different lengths (1 metre, 0.5 metre and 2 metres) activated between metre 12 and metre 24 of FA4.

An experiment was conducted at Vattenfall R&D facilities to assess the feasibility of using these hybrid cables to measure fast changes in water velocity in the range of 5-20 cm/s. A 35-metre section of FA4 was deployed in a test water flume. Two equal and parallel 17-m sections of FA4, separated by less than 10 centimetres, were laid at the bottom part of the flume by bending the whole length of FA4 on itself as shown in Fig. 7(a). In first 17-m section, the metal wires were contacted between position 6m and 9m to create a 3-metre hotspot. The other half was left unheated and used as reference fibre.

The height of the water level was changed several times at a fixed flow rate of to 100 L/min to obtain different flow speeds in the water flume. A flowmeter was used to monitor the water speed at one position in the flume. A similar power rate of around 20W/m was supplied to the hotspot. Constant heating and pulsed heating at the rate of 1 min ON and 2 minutes OFF were both tested during the measurement campaign.

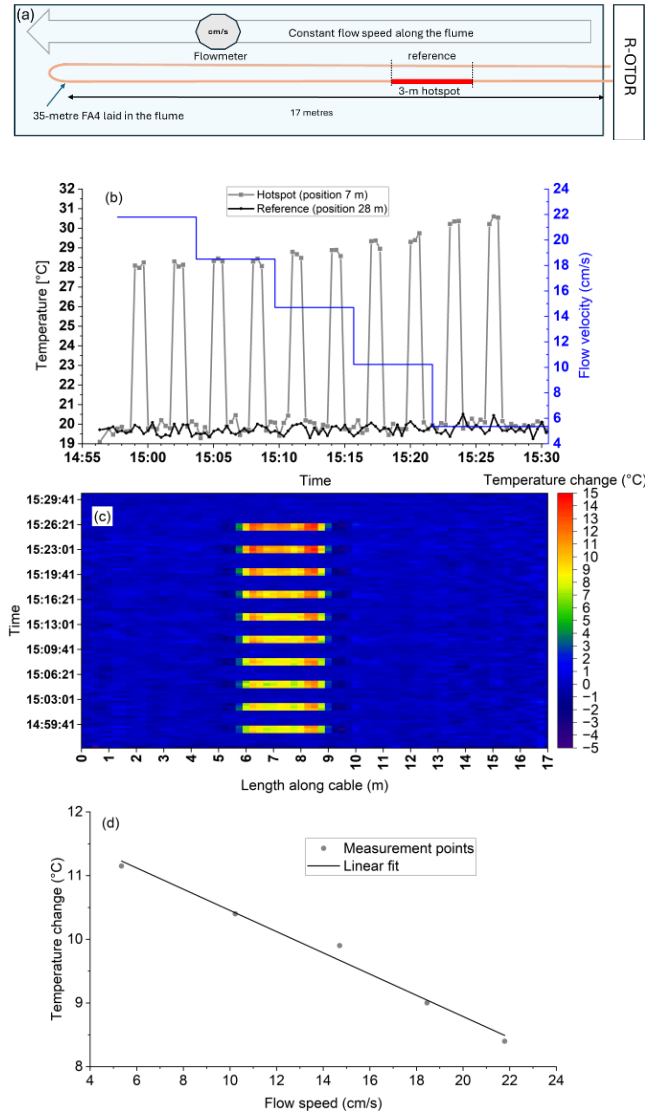

Fig. 7. (a) Schematic representation of the test perfomed in a water flume. (b) Measured temperature of the hotspot and (c) heatmap of the distributed temperature change along the first 17-m of FA1 at different water speeds. (d) Temperature change (from no heating) v/s flow speed.

The distributed temperature along the 35-metre FA4 cable was measured every 20 seconds with a 1 metre spatial resolution using a commercial Raman-OTDR in dual-ended configuration. Fig. 7(b) shows the temperature evolution measured at the hotspot (position 7 metres on FA4) and on the reference part (position 28 metres on FA4) of the cable.
The corresponding distributed temperature difference, directly linked to the change in heat dissipation due to varying flow speed, is obtained by direct point by point comparison of the distributed temperature measurements along the two parallel lengths of heated and unheated cables (Fig. 7(c)). It is clear on Fig. 7(d) that a change in temperature at the hotspot location can be easily measured with varying water flow speed (21.8 to 5.3 cm/s). Hence, these hybrid assemblies can represent a cheaper, energy-effective alternative to thick optical cables commonly deployed for active distributed temperature sensing for applications where robust cables are not required.

## V. OUTLOOK AND FUTURE WORK

Fibre-optic assemblies bring another approach to fibre-optic sensor integration by facilitating their deployment and use in assets, while allowing for multi-parameter sensing using a single thin fibre-optic cable. These assemblies can be made as thin as 100μm with multi-parameter sensor capabilities by using ultra-thin fibres if seamless embedding in composite is required. They can be produced over kilometers, enabling the use of distributed fibre sensing techniques with these assemblies.

This modular approach makes the design options for these miniature cables almost endless, with free choice of the number of optical fibres, their types (single-mode, multimode, Fibre Bragg Grating arrays, etc.,), their dimensions, and coating material. In this communication, we presented fibre-optic assemblies made up of typical optical fibres used for sensing. By associating them in the same fibre-optic assembly, multi-parameter sensing could be achieved using different interrogation systems as described in Use Case 1.

Co-extruding several single mode fibres or FBG arrays can also help unlock the measurement of new parameters such as distributed curvature and shape. The separation between the optical cores can be chosen to be much larger than 35 μm, the typical value in multicore fibres, thereby allowing curvature measurements with higher sensitivity with a simpler and more cost-effective system as in Use Case 2. Single-mode and multimode fibres can also be combined with more specialized fibres, such as large mode area fibres for power delivery and hollow-core fibres in the same assembly to bring in added functionalities on top of fibre-optic sensing.

The buffer material is an important part of the fibre-optic assembly design and can be selected from a range of standard fibre-optic buffering materials, such as Nylon, ETFE, PVC, Hytrel, PEEK, PFA, etc., to reduce micro-bend losses during embedding or to bring additional protection to the optical fibres to harsh environment or hazardous products during deployment and usage. Distributed monitoring of other parameters such as humidity and chemical species can be unlocked through astute choice of coating and buffer materials on these fibre assemblies to allow the latter to be permeable to specific gases and liquids that need to be monitored while assuring robust integration in assets.

Thin electrical wires can also be added as part of the assembly, allowing the use of this hybrid fibre-optic assembly for active DTS measurements for by heating desired sections of the optical fibre by Joule effect through the wires. This can be exploited for flow speed measurements as described in Use Case 3. The measurement of electrical parameters (resistivity, current, capacitance) of the electrical wires at specific locations where the metal is directly exposed to the material adds another sensing dimension to the hybrid assembly since new parameters, that cannot be monitored by light, can be integrated in the system. These electrical measurements can also be associated to distributed fibre-optic sensing techniques for multi-parameter and multi-sensing approaches using the same optical fibre. For instance, they can be used to measure temperature on the same optical fibre section for data reliability, especially in harsh conditions where one method could fail or give low signal to noise ratio or used to discriminate between measurands while performing multi-parameter sensing with the same fibre-optic assembly.

The use cases reported in this paper show the critical need to carefully consider fibre-optic cable diameter, with robustness when designing application-specific sensing solutions. Thus, the proposed modular fibre-optic assembly concept offers countless possibilities to the optical fibre sensing community.